\RequirePackage{fix-cm}
\documentclass[twocolumn,epjc3]{svjour3}  
\smartqed  
\RequirePackage{graphicx}
\usepackage{amsmath}
\journalname{Eur. Phys. J. C}
\begin{document}

\title{Heavy baryon decay widths in the large $N_c$ limit \\ in chiral theory 
}


\author{Michal Praszalowicz\thanksref{e1,addr1}
      }

\thankstext{e1}{e-mail: michal@if.uj.edu.pl}


\institute{M. Smoluchowski Institute of Physics, Jagiellonian University,  S. {\L}ojasiewicza 11,
30-348 Krak{\'o}w, Poland \label{addr1}
}

\date{Received: date / Accepted: date}

\maketitle

\begin{abstract}
We propose large $N_c$ generalizations for the "diquark" representations of SU(3)$_{\rm flav}$
relevant for  positive parity heavy baryons,
including putative exotic states. Next, within the framework of the Chiral Quark Soliton Model, we calculate
heavy baryon  masses and decay widths. We show that in the limit of $N_c \rightarrow \infty$ {\em all} decay widths vanish, 
including the widths of exotica. This result is in fact more general than the model itself, as it relies only on the
underlying symmetries: {\em i.e.} SU(3)$_{\rm flav}$ and hedgehog symmetry. Furthermore, using explicit
model formulae for the decay constants in the non-realtivistic limit, we show that there is a hierarchy of the decay 
couplings, which may explain observed pattern of experimental widths.

\keywords{Chiral symmetry \and Hedgehog symmetry \and Heavy baryons \and Large $N_c$ limit}
\end{abstract}

\section{Introduction}
\label{intro}

Recently the LHCb Collaboration at CERN announced a discovery of five narrow $\Omega^0_c$ resonances 
with masses ranging from 3 to 3.2 GeV~\cite{Aaij:2017nav}, that have been later confirmed by BELLE~\cite{Yelton:2017qxg}.
The widths of these resonances is of the order of a few MeV, with two of them being exceedingly small: 
$\Gamma(\Omega_c^0(3050))=0.8\pm 0.2 \pm 0.1$  and 
$\Gamma(\Omega_c^0(3119))=1.1 \pm 0.8 \pm 0.4$~MeV. 
In Refs.~\cite{Kim:2017jpx,Kim:2017khv} we have proposed to interpret these two narrow states
as exotic pentaquarks using as a guidance the Chiral Quark Soliton Model~\cite{Diakonov:1987ty} 
($\chi$QSM -- for review see Refs.~\cite{Christov:1995vm,Alkofer:1994ph}). Other possible interpretations
of these sta{\-}tes are summarised in Ref.~\cite{Praszalowicz:2018uoa}.
The situation here is similar to 
the light pentaquark state $\Theta^+$~\cite{Praszalowicz:2003ik,Diakonov:1997mm}, 
which -- if it exists -- has to be very narrow. Indeed, the evidence for $\Theta^+$ that survived until
now after the first announcement in 2003~\cite{Nakano:2003qx,Barmin:2003vv} is the analysis 
by DIANA Collaboration~\cite{Barmin:2013lva}
that requires $\Gamma_{\Theta^+} \sim 0.3$~MeV (see also \cite{Nakano:2017fui}). On theoretical side it has been shown
in Ref.~\cite{Diakonov:1997mm} that in the non-relativistic limit of the $\chi$QSM the relevant decay coupling of the exotic antidecuplet
vanishes identically. This might explain the required smallness of $\Theta^+$ decay width.

The nullification of the pertinent decay coupling in the non-relativistic limit  occurs only if 
the rotational sub-leading  $1/N_c$ contributions
are taken into account \cite{Diakonov:1997mm}. 
It has been subsequently shown in Ref.~\cite{Praszalowicz:2003tc} that the cancellation of terms that are
of different order in $N_c$ is consistent with the large $N_c$ limit if the baryon $SU(3)_{\rm flav}$ representations
are appropriately enlarged to account for colour neutrality. 
So despite the fact that formally $\Gamma_{\Theta^+}(N_c \rightarrow \infty)={\cal{O}}(1)$ (while the decuplet decay width
$\Gamma_{\Delta}(N_c \rightarrow \infty)={\cal{O}}(1/N_c^2)$) the smallness of the decay width is assured 
by another small parameter 
(that, however, has not been analytically defined) related to degree of "relativisticity".

In Ref.~\cite{Kim:2017jpx,Kim:2017khv,Yang:2016qdz} a phenomenological  
analysis of heavy baryon properties has been performed 
in the framework of the $\chi$QSM (see also \cite{Kim:2018xlc,Yang:2018uoj,Kim:2018nqf}). 
It turned out that all decay widths have been very well reproduced \cite{Kim:2017khv}, 
also the two narrowest
ones of the putative pentaquarks.
In the present paper we want to find out whether a suppression
mechanism similar to the one discussed above
could explain extraordinary small widths of two narrowest $\Omega_c^0$ states reported by the LHCb
(given their interpretation as exotica),
or whether the smallness of these widths is a pure numerical coincidence. 

In the present paper, extending  Ref.~ \cite{Kim:2017khv}, we present an analysis, which
shows that there is a hierarchy of the decay constants that indeed suppresses decay widths
of heavy pentaquark states, and that degree of this suppression depends on the decay channel. While
this result has been to some extent expected from our experience with light quark exotica, the other result
that {\em all} decay widths of heavy baryons studied here vanish in the large $N_c$ limit (in contrast
to the case of $\Theta^+$), even if we do not take the non-relativistic limit, comes as a surprise.

The paper is organised as follows. In the next Section we briefly recapitulate main features of the
$\chi$QSM. Then, in Sect.~\ref{sec:reps}, we show how SU(3)$_{\rm flav}$ representations for 
the light subsystem in heavy baryons have to be generalised to the case of  $N_c >3$. This prescription
is used in the Appendix to provide the relevant Clebsch-Gordan coefficients needed to compute the decay
widths in Sect.~\ref{sec:dw}. To calculate the widths we need mass formulae to calculate the momentum of
the outgoing meson, what is done in Sect.~\ref{sec:ms}. We summarise in Sect.~\ref{sec:sum}.

\section{Chiral Quark Soliton Model\\ for heavy baryons}
\label{sec:cdsm4HB}

The $\chi$QSM is based on an argument of Witten
\cite{Witten:1979kh,Witten2,Witten3} that in the limit of large number of colors,
$N_{c}$ relativistic valence quarks generate chiral mean fields
represented by a distortion of a Dirac sea that in turn 
influence the valence quarks themselves  forming a self-organised configuration called a
\emph{soliton}.  The soliton configuration corresponds to the solution of the Dirac equation for the constituent quarks (with
gluons integrated out) in the mean-field approximation where the mean fields respect so called {\em hedgehog}
symmetry. Since it is impossible to construct a pseudoscalar field that changes sign under
inversion of coordinates, which would be compatible with
the SU(3)$_{\rm flav}\times$SO(3) space symmetry, one has to resort to 
a smaller {\em hedgehog}  symmetry that, however,
leads to the correct baryon spectrum.

Next, rotations of
the soliton, both in flavor and configuration spaces, are quantised
semiclassically and the collective Hamiltonian  is computed. The model
predicts rotational 
baryon spectra that satisfy the following selection rules:
\begin{itemize}
\item allowed SU(3) representations must contain states with hypercharge
$Y^{\prime}=N_{\rm val}/3$,

\item the isospin ${ T}^{\prime}$ of the states with $Y^{\prime}%
=N_{\rm val}/3$ is equal the soliton spin ${ J}$
\end{itemize}
where $N_{\rm val}$ denotes the number of valence quarks.

Rotational energy reads as follows \cite{Guadagnini:1983uv,Mazur:1984yf,Jain:1984gp}:
\begin{eqnarray}
\mathcal{E}_{\left(p,\, q\right)}^{\mathrm{rot}} 
&= &  M_{\mathrm{sol}} 
    + \frac{J(J+1)}{2I_{1}}  
\cr
&  & +\;\frac{C_2(p,q)-J(J+1) -3/4\, Y^{\prime\, 2}}{2I_{2}}
\label{eq:HE}
\end{eqnarray}
where $C_2$ denotes SU(3) Casimir operator and $J$ stands for the soliton spin. Soliton mass
$M_{\mathrm{sol}}$ and moments of inertia $I_{1,2}$ are calculable in terms of relativistic single quark
wave functions.

For light baryons $N_{\rm val}=N_c$ and the lowest SU(3)$_{\rm flav}$ representations allowed by the above
selection rules are octet of spin 1/2, decuplet of spin 3/2 and exotic anti-decuplet of spin 1/2. $M_{\mathrm{sol}}$ and  $I_{1,2}$
scale like $N_{\rm val}$.

Recently we have proposed~\cite{Yang:2016qdz}, following Ref.~\cite{Diakonov:2010zz}, how
to generalise the above approach to heavy baryons,
by stripping off one valence quark 
and replacing it by a heavy quark to neutralise the color. 
In the large $N_c$ limit both systems: light and heavy baryons are described essentially by the same mean field, and the
only difference is now that $N_{\rm val}=N_c-1$. The lowest allowed SU(3) representations are in this case (as in the quark model)
 $\overline{\mathbf{3}}$ of spin 0  and 
to ${\mathbf{6}}$ of spin 1. Therefore, the baryons constructed  from such a soliton and a heavy
quark form an SU(3) anti-triplet of spin 1/2 and two sextets  of spin 1/2 and 3/2 that are subject to a hyper-fine
splitting. The first exotic representation is  $\overline{\bf 15}$ with spin 0 or 1. 
 However, as can be seen from Eq.~(\ref{eq:HE}), the spin 1 soliton is
lighter\footnote{Explicit calculations and phenomenological fits show that $1/I_1<1/I_2$.}, 
hence in the following we ignore the one with spin 0. This means that exotic heavy pentaquarks belonging 
to the SU(3)$_{\rm flav}$ $\overline{\bf 15}$ have total spin 1/2 and 3/2. These multiplets are hyperfine split 
with splitting parameter proportional to $1/m_Q$.

\section{Large $N_c$ representations for heavy baryons }
\label{sec:reps}

\begin{figure}
\begin{centering}
  \includegraphics[width=0.45\textwidth]{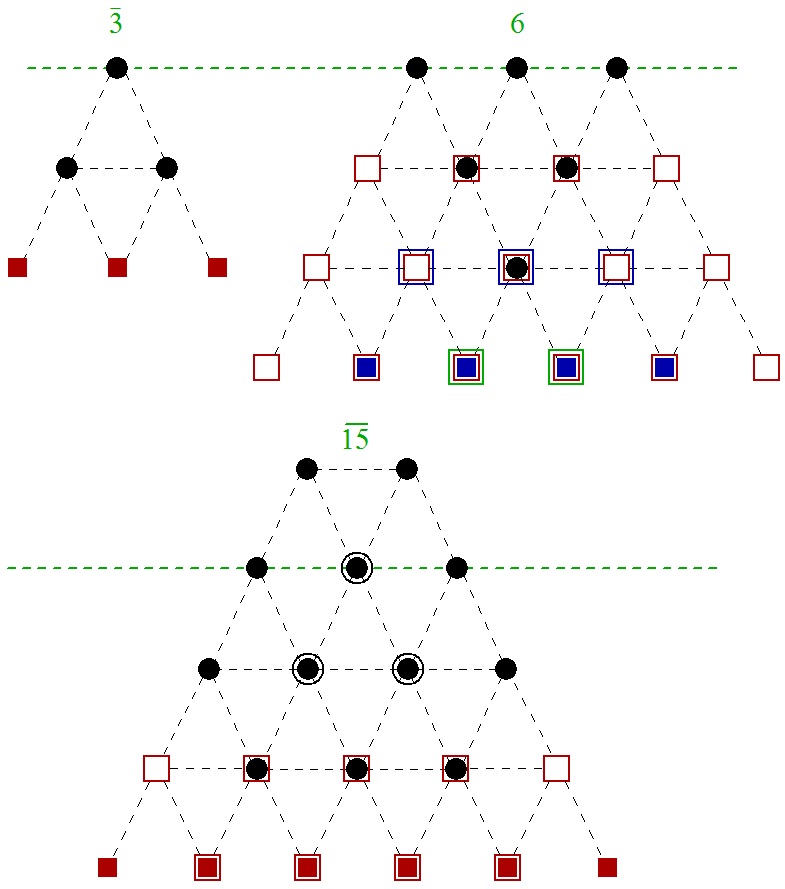}
 \end{centering}
\caption{Large $N_c$ generalizations of weigth diagrams of SU(3)$_{\rm flav}$ representations 
$\bar{\bf 3}$, ${\bf 6}$ and $\overline{\bf 15}$. Black circles denote physical 
states that exist for $N_c=3$. Squares denote spurious states that disappear 
for $N_c=3$.  It is understood that these diagrams continue towards negative values of $Y$.
Horizontal dashed (green) lines correspond to $Y'=(N_c-1)/3$. }
\label{fig:reps}       
\end{figure}

For $N_c>3$ we have to generalise $\bar{\bf 3}=(0,1)$\footnote{We use here another notation for SU(3) representation
expressed in terms of $p$ quark indices and $q$ anti-quark indices: $(p,q)$.}, ${\bf 6}=(2,0)$ and $\overline{\bf 15}=(1,2)$
to the case of arbitrary (odd) $N_c$~\cite{Karl:1985qy,Bijnens:1985ry,Bijnens2,Dulinski:1987er}.
In this case the $\chi$QSM constraint generalises to $Y'=(N_c-1)/3$. This criterion has to be supplemented
by yet another condition, which is usually a requirement that large $N_c$ solitons (and therefore baryons) have the same
spin as in the $N_c=3$ case. This means that the pertinent  representations have the same number of quark indices $p=p_0$
as for  $N_c=3$, but different $q$. In the quark model language this corresponds to the addition of an antisymmetrised quark pair to
a given baryon wave function when we increase $N_c$ by 2. This means that the number of antiquark indices  $q_0$ 
at $N_c=3$ has to be replaced by $q_0+(N_c-3)/2$. Therefore we arrive at the following generalisations:
\begin{eqnarray}
"{\overline{\bf3}}"  &  = &(0,1+q),\;\dim("{\overline{\bf 3}%
}")=\frac{1}{2}(2+q)(3+q),\nonumber\\
"{\bf 6}"  &  =&(2,q),\;\;\;\;\;\;\; \dim("{\bf 6}")=\frac{3}{2}%
(1+q)(4+q),\nonumber\\
"\overline{\bf{15}}"  &  =&(1,2+q),\; \dim("\overline{\bf{15}%
}")=(3+q)(5+q)
\label{eq:reps}
\end{eqnarray}
with%
\begin{equation}
q=\frac{N_{c}-3}{2} \label{eq:q}%
\end{equation}
that are illustrated in  Fig.~\ref{fig:reps}.

%

It is now clear that various matrix elements of the irreducible  SU(3)$_{\rm flav}$
tensor operators will acquire $N_c$ dependence if sandwiched between states
belonging to representations (\ref{eq:reps}). In this respect there is no difference
between the quark model and the $\chi$QSM.  Indeed, it possible to show on general 
grounds that representation content of the quark model and soliton
model coincide for large $N_c$ \cite{Manohar:1984ys,Jenkins:2004tm}. 
The difference appears because due to
the {\rm hedgehog} symmetry the $\chi$QSM provides certain relations between reduced
matrix elements in different multiplets, which in the naive quark model are arbitrary.  

\bigskip

\section{Heavy baryon masses\\ in the 
Chiral Quark Soliton Model}
\label{sec:ms}

In the $\chi$QSM the soliton is quantised as a symmetric top and the pertinent mass
formula for heavy baryons takes the following form:
\begin{equation}
M_B=m_Q+\mathcal{E}_{\left(p,\, q\right)}^{\mathrm{rot}} + \delta_B +\Delta_B^{\rm hf}
\label{eq:mass}
\end{equation}
where $m_Q$ stands for the heavy quark mass. Rotational soliton energy is given by (\ref{eq:HE}) and
mass splittings due to the non-zero strange quark mass $m_s$ are denoted by $\delta_B$, and 
$\Delta_B^{\rm hf}$ denotes hyperfine splitting which vanishes in a heavy quark limit. These two
contributions are not important for the discussion of the large $N_c$ limit.

Mass differences of heavy baryon multiplets are therefore equal to differences of rotational energies:
\begin{eqnarray}
\mathcal{E}_{\bf 6}^{\mathrm{rot}} - \mathcal{E}_{\bar{\bf 3}}^{\mathrm{rot}} &=& \frac{1}{I_1} \sim \frac{1}{N_c}, \nonumber \\
\mathcal{E}_{\overline{\bf 15}}^{\mathrm{rot}} - \mathcal{E}_{\bar{\bf 3}}^{\mathrm{rot}} &=& \frac{N_c+1}{4 I_2}+\frac{1}{I_1} \sim
N_c^0, \nonumber \\
\mathcal{E}_{\overline{\bf 15}}^{\mathrm{rot}} - \mathcal{E}_{{\bf 6}}^{\mathrm{rot}} &=& \frac{N_c+1}{4 I_2} \sim
N_c^0.
\label{eq:Erotdiff}
\end{eqnarray}
We see from Eq.~(\ref{eq:Erotdiff}) that regular multiplets are degenerate in the large $N_c$ limit, whereas
the exotic multiplet, namely $\overline{\bf 15}$, remains heavier by ${\mathcal O}(1)$. Here the situation
is identical as in the case of light baryons, where the mass difference between decuplet and octet vanishes for
$N_c \rightarrow \infty$, while splitting to the exotic anti-decuplet does not. This behaviour results in
the non-vanishing decay width of the exotic $\overline{\bf 10}$, which was the main argument against the consistency
of the $\chi$QSM to light baryon exotica \cite{Pobylitsa:2003ju,Cherman:2004qx}. 
We will see in the following that, despite (\ref{eq:Erotdiff}),
decay widths of exotic heavy baryons do vanish for large $N_c$.

\section{Decay widths}
\label{sec:dw}

The $\chi$QSM allows to compute strong decay wid{\-}ths 
that proceed by the soliton transition to another
configuration with emission of  a pseudoscalar meson $\varphi$. 
In the present paper  following \cite{Kim:2017khv} we use strong decay widths of
nonexotic and exotic heavy quark baryons (both charm and bottom) computed in an approach proposed many years ago
by Adkins, Nappi and Witten \cite{Adkins:1983ya} and expanded in Ref.~\cite{Diakonov:1997mm}, 
which is based on the Goldberger-Treiman
relation where strong decay constants are expressed in terms of the axial current couplings
(see Ref.~\cite{Yan:1992gz} for the derivation in the case of heavy baryons).
In this case
the decay operator can be expressed in terms of the weak axial decay constants\footnote{
For reader's convenience we give the relations of the constants
$a_{1,2,3}$ to nucleon axial charges in the chiral limit:  
$g_{A}=\frac{7}{30}\left(
  -a_{1}+\frac{1}{2}a_{2}+\frac{1}{14}a_{3}\right)$, $g_A^{(0)}=\frac
12 a_3$,  
$g_{A}^{(8)}=\frac{1}{10 \sqrt 3}\left(
  -a_{1}+\frac{1}{2}a_{2}+\frac{1}{2}a_{3}\right)$.}  $a_i$ and meson decay constant $F_{\varphi}$:
\begin{eqnarray}
\mathcal{O}^{(8)}_{\varphi}&=& \\
&-&\left[  a_{1}D_{\varphi\,i}^{(8)}+a_{2}\,d_{ibc}%
D_{\varphi\,b}^{(8)}\hat{J}_{c}+a_{3}\frac{1}{\sqrt{3}}D_{\varphi\,8}%
^{(8)}\hat{J}_{i}\right] 
 \frac{p_{i}}{2F_{\varphi}} \nonumber
 \label{eq:dec-op}%
\end{eqnarray}
where $p_{i}$ is the
c.m. momentum of the outgoing meson of mass 
$m$:%
\begin{eqnarray}
\left\vert p_i \right\vert &=& p \\
&=& \frac{\sqrt{(M_{1}^{2}%
-(M_{2}+m)^{2})(M_{1}^{2}-(M_{2}-m)^{2})}
}{2M_{1}}. \nonumber 
\end{eqnarray}
It is important to note that in the chiral limit where $m \rightarrow 0$ 
momentum $p$ behaves differently with $N_c$ , due to
(\ref{eq:Erotdiff}), depending on the initial
and final flavor representations:
\begin{equation}
p_{ {\bf 6} \rightarrow \overline{\bf 3}} \sim \frac{1}{N_c}, \;\; p_{\overline{\bf 15} \rightarrow {\bf 6}, \overline{\bf 3}} \sim N_c^0.
\label{eq:pNc}
\end{equation}
This $N_c$ counting is of primary importance for correct determination of the $N_c$ dependence of the decay
widths.

The decay width for $B_1 \rightarrow B_2 + \varphi$  is  related to the matrix element of
$\mathcal{O}^{(8)}_{\varphi }$ squared, summed over the final and averaged
over the initial spin and isospin  denoted as $\overline{\langle 
\ldots \rangle^{2}}$,  see the Appendix of
Ref.~\cite{Diakonov:1997mm} for details of the corresponding
calculations:
\begin{equation}
\Gamma_{B_{1}\rightarrow B_{2}+\varphi}=\frac{1}{2\pi}\overline{\left\langle
B_{2}\left\vert \mathcal{O}_{\varphi}\right\vert B_{1}\right\rangle ^{2}%
}\,\frac{M_{2}}{M_{1}}p.
\end{equation}
  Factor $M_{2}/M_{1}$ follows from the heavy baryon chiral
perturbation theory, see {\em e.g.} Ref.~\cite{Cheng:2006dk,Cheng2}. 
While it is important for phenomenological applications, it is 
irrelevant for our discussion as it scales like $N_c^0$.

The final formula for the decay width in terms of axial constants $a_{1,2,3}$ reads as follows :%
\begin{eqnarray}
\label{eq:Gammageneral}
\Gamma_{B_{1}\rightarrow B_{2}+\varphi}  &  =& \frac{1}{24 \pi}\frac{p^{3}%
}{F_{\varphi}^{2}}\frac{M_{2}}{M_{1}}G_{\mathcal{R}_{1}\rightarrow
\mathcal{R}_{2}}^{2} \frac{\dim\mathcal{R}_{2}}{\dim\mathcal{R}_{1}}\\
&  &\left[
\begin{array}
[c]{cc}%
{\bf{8}} & \mathcal{R}_{2}\\
01 & Y^{\prime}S_{2}%
\end{array}
\right\vert \left.
\begin{array}
[c]{c}%
\mathcal{R}_{1}\\
Y^{\prime}S_{1}%
\end{array}
\right]  ^{2}\left[
\begin{array}
[c]{cc}%
{\bf{8}} & \mathcal{R}_{2}\\
Y_{\varphi}T_{\varphi} & Y_{2}T_{2}%
\end{array}
\right\vert \left.
\begin{array}
[c]{c}%
\mathcal{R}_{1}\\
Y_{1}T_{1}%
\end{array}
\right]  ^{2}\nonumber
\end{eqnarray}
Here $\mathcal{R}_{1,2}$ are the SU(3) representations of the initial and final baryons
and $[.. .|..]$ are SU(3) iso-scalar factors.
The decay constants $G_{\mathcal{R}_{1}\rightarrow
\mathcal{R}_{2}}$ are calculated from the matrix elements of
(\ref{eq:dec-op}) for representations (\ref{eq:reps}) and read as follows: %
\begin{eqnarray}
G_{{\bf 6} \rightarrow \overline{\bf 3}}=H_{{\overline{\bf{3}}}}  & =&-\tilde{a}_{1}+\frac{1}{2}a_{2}%
,\nonumber\\
G_{\overline{\bf 15} \rightarrow \overline{\bf 3}} = G_{\bf{\overline{3}}}  & =&-\tilde{a}_{1}-\frac{N_{c}-1}{4}%
a_{2},\nonumber\\
G_{\overline{\bf 15} \rightarrow {\bf 6}} = G_{\bf 6}  & =&-\tilde{a}_{1}-\frac{N_{c}-1}{4}a_{2}-a_{3}.
\label{eq:Gammagen}
\end{eqnarray}

In the $\chi$QSM one can define so called non-relativistic (or quark model QM) limit 
\cite{Diakonov:1997mm,Praszalowicz:1995vi,Praszalowicz:1998jm}
by squeezing 
the soliton to zero. The easiest way to perform this limit is to use the variational approach,
in which one solves the Dirac equation for single quark energy levels in the {\em hedgehog}
mean field characterised by a variational parameter $r_0$, which is called the soliton size.
For the physical solution the value of $r_0$ is determined by
the balance of the valence quark contribution that decreases with $r_0$ and the contribution
of the appropriately regularised Dirac sea that increases with $r_0$. The QM limit is defined by 
taking {\em artificially} $r_0 \rightarrow 0$.
In this limit  the valence level reaches its free energy value equal to the constituent  mass $M$. At the same
time the contribution of the Dirac sea is approaching zero\footnote{This justifies the name: Quark Model limit,
because the soliton energy is equal essentially to $N_{\rm val} \times M$.}, since the soliton energy is evaluated with
respect to the unperturbed Dirac sea. In the QM limit  parameters $a_i$ can be computed analytically 
\cite{Praszalowicz:1995vi,Praszalowicz:1998jm}.
One has to observe that in the present case the number of valence quarks is $N_c-1$ rather than $N_c$,
and therefore the only $N_c$ dependent parameter $a_1$ has to be appropriately  rescaled; that is why we
have used a "$\tilde{~}$" over $a_1$ \cite{Kim:2017khv}. We have \cite{Praszalowicz:1995vi,Praszalowicz:1998jm}:
\begin{equation}
-\tilde{a}_{1}\stackrel{\rm QM}{\rightarrow} N_{c}+1,\;\;
a_{2}\stackrel{\rm QM}{\rightarrow}4,\;\;
a_{3}\stackrel{\rm QM}{\rightarrow} 2
\end{equation}
and we get a hierarchy between the decay constants in the QM limit: 
\begin{equation}
H_{{\overline{\bf{3}}}}\stackrel{\rm QM}{\rightarrow}N_c+3,  \; \;
G_{\bf{\overline{3}}}\stackrel{\rm QM}{\rightarrow}2, \; \;
G_{\bf 6}\stackrel{\rm QM}{\rightarrow} 0.
\label{eq:HGQM}
\end{equation}

By this observation we have argued in Ref.~\cite{Kim:2017khv} that the decays of exotic 
$\Omega_c^0$ resonances should be suppressed with respect to the decays
of regular baryons that are driven by the unsuppressed constant $H_{{\overline{\bf{3}}}}$.

However, even off the QM  limit, where all couplings 
$H_{{\overline{\bf{3}}}}, \;G_{\bf{\overline{3}}},\;  G_{\bf 6} \sim N_c$,
decays of the exotic $\Omega_c$'s are suppressed
due to the $N_c$ dependence of the pertinent iso{\-}sca{\-}lar factors in Eq.~ (\ref{eq:Gammageneral}).
Indeed, for the energetically  allowed decays we have:%
\begin{eqnarray}
\Gamma_{\Sigma({\bf{6}}_{1})\rightarrow\Lambda(\overline{\bf{3}%
}_{0})+\pi} &  =&\frac{1}{72\pi}\frac{M_{\Lambda(\overline{\bf{3}}%
_{0})}}{M_{\Sigma({\bf{6}}_{1})}}\frac{p^{3}}{F_{\pi}^{2}}\nonumber \\
&\times&\frac
{(N_{c}-1)(N_{c}+3)}{(N_{c}+1)(N_{c}+5)}H_{{\overline{\bf{3}}}}%
^{2},\nonumber
\\
\Gamma_{\Xi({\bf{6}}_{1})\rightarrow\Xi(\overline{\bf{3}}%
_{0})+\pi} &  =&\frac{1}{72\pi}\frac{M_{\Xi(\overline{\bf{3}}_{0})}%
}{M_{\Xi({\bf{6}}_{1})}}\frac{p^{3}}{F_{\pi}^{2}} \nonumber \\
&\times&\frac{N_{c}^{2}}%
{(N_{c}+1)(N_{c}+5)}H_{{\overline{\bf{3}}}}^{2},\nonumber\\
\Gamma_{\Omega(\overline{\bf{15}}_{1})\rightarrow\Xi(\overline
{\bf{3}}_{0})+K} &  =&\frac{4}{3\pi}\frac{M_{\Xi(\overline
{\bf{3}}_{0})}}{M_{\Omega(\overline{\bf{15}}_{1})}}\frac
{p^{3}}{F_{K}^{2}}\nonumber \\
&\times & \frac{G_{{\overline{\bf{3}}}}^{2}}{(N_{c}%
+1)(N_{c}+5)(N_{c}+7)},\nonumber
\end{eqnarray}
\begin{eqnarray}
\Gamma_{\Omega(\overline{\bf{15}}_{1})\rightarrow\Omega({\bf{6}}%
_{1})+\pi} &  =&\frac{4}{27\pi}\frac{M_{\Omega({\bf{6}}_{1})}}%
{M_{\Omega(\overline{\bf{15}}_{1})}}\frac{p^{3}}{F_{\pi}^{2}} \nonumber \\%
&\times& \frac{G_{\bf{6}}^{2}}{(N_{c}+1)(N_{c}+7)}\gamma,\nonumber\\
\Gamma_{\Omega(\overline{\bf{15}}_{1})\rightarrow\Xi({\bf{6}}%
_{1})+K} &  =&\frac{8}{27\pi}\frac{M_{\Xi({\bf{6}}_{1})}}{M_{\Omega
(\overline{\bf{15}}_{1})}}\frac{p^{3}}{F_{K}^{2}}\nonumber \\
&\times& \frac
{G_{\bf{6}}^{2}}{(N_{c}+1)^{2}(N_{c}+7)}\gamma.
\label{eq:decw}
\end{eqnarray}
For multiplets where the soliton spin $J$ (denoted by a subscript at the representation label in Eq.~(\ref{eq:decw}))
is equal to one, hyperfine splittings
to a heavy quark result in  two spin multiplets  1/2 and 3/2.
Factors $\gamma$ take this additional couplings into account\footnote{see Eratum in Ref.~\cite{Kim:2017khv}.}:
\begin{eqnarray}
\gamma(1/2\rightarrow1/2)={2}/{3},\qquad & & \gamma(1/2\rightarrow3/2)={1}%
/{3},\nonumber\\
\gamma(3/2\rightarrow1/2)={1}/{6},\qquad &  &\gamma(3/2\rightarrow3/2)={5}/{6}.
\end{eqnarray}%

Armed with explicit formulae for the decay widths (\ref{eq:decw}), for the pertinent
couplings (\ref{eq:Gammagen}), for $N_c$ meson momentum dependence (\ref{eq:pNc}),
and remembering that $F_{\varphi}^2 \sim N_c$, we can compute $N_c$ dependence
of the decay widths and, using (\ref{eq:HGQM}), $N_c$ dependence
of the decay widths in the Quark Model limit: 
\begin{eqnarray}
  \Gamma_{\Sigma({\bf{6}}_{1})\rightarrow\Lambda(\overline
{\bf{3}}_{0})+\pi}&\stackrel{N_c \rightarrow \infty}{\rightarrow}&%
\frac{1}{N_{c}^{2}}\stackrel{\rm QM}{\rightarrow}\frac{1}{N_{c}^{2}%
},\nonumber\\
\Gamma_{\Xi({\bf{6}}_{1})\rightarrow\Xi(\overline{\bf{3}}%
_{0})+\pi}&\stackrel{N_c \rightarrow \infty}{\rightarrow}&\frac{1}{N_{c}^{2}%
}\stackrel{\rm QM}{\rightarrow}\frac{1}{N_{c}^{2}},\nonumber\\
\Gamma_{\Omega(\overline{\bf{15}}_{1})\rightarrow\Xi(\overline
{\bf{3}}_{0})+K}&\stackrel{N_c \rightarrow \infty}{\rightarrow}& \frac
{1}{N_{c}^{2}}\stackrel{\rm QM}{\rightarrow} \frac{1}{N_{c}^{4}},\nonumber\\
 \Gamma_{\Omega(\overline{\bf{15}}_{1})\rightarrow\Omega
({\bf{6}}_{1})+\pi}&\stackrel{N_c \rightarrow \infty}{\rightarrow}&%
\frac{1}{N_{c}}\stackrel{\rm QM}{\rightarrow} 0,\nonumber\\
\Gamma_{\Omega(\overline{\bf{15}}_{1})\rightarrow\Xi({\bf{6}}%
_{1})+K}&\stackrel{N_c \rightarrow \infty}{\rightarrow}&\frac{1}{N_{c}^{2}%
}\stackrel{\rm QM}{\rightarrow} 0.
\label{eq:gamlimit}
\end{eqnarray}

Equations (\ref{eq:gamlimit})  show that {\em all} widths relevant for heavy
baryon decays, including exotica, vanish for $N_c \rightarrow \infty$. This
result is quite obvious for regular baryons that are degenerate in this limit
 (see Eqs.~(\ref{eq:Erotdiff}), (\ref{eq:pNc})), and
the quadratic dependence on $1/N_c$ is the same as in the case of {\em e.g.}
$\Delta$ decay. It is however surprising that for $N_c \rightarrow \infty$ 
exotic states that are not degenerate
with the ground state heavy baryons  
 (see again Eqs.~(\ref{eq:Erotdiff}) and (\ref{eq:pNc})),
have nevertheless decay
widths that tend to zero in contrast with the decay width of the putative light
pentaquark $\Theta^+$. For a decay  linking  baryons of the same isospin 
the suppression power
is weaker by one. In the Quark Model limit the decay widths
of exotica are, however, further suppressed.
This is an interesting situation not known from the light baryons and it deserves
more detailed studies.

\section{Summary}
\label{sec:sum}

Prompted by the pentaquark assignment of two narrowest $\Omega_c^0$ states reported recently by the
LHCb Collaboration we have studied the large $N_c$ limit of the decay widths of heavy quark baryons
within the Chiral Quark Soliton Model. We have calculated all energetically allowed strong decays of
the ground state SU(3)$_{\rm flav}$ sextet and of the putative pentaquark $\Omega_c^0$'s. To this end we have used
heavy baryon chiral perturbation theory and the Glodberger-Treiman relation for heavy baryons. 

We have proposed a natural enlargement of the pertinent SU(3)$_{\rm flav}$ representations for $N_c \rightarrow \infty$
and calculated the relevant matrix elements obtaining analytical results for arbitrary (odd) $N_c$. This required to calculate
SU(3) Clebsch-Gordan coefficients for large representations (\ref{eq:reps}). 
The relevant technique has been briefly discussed in the Appendix.

The main result is that {\em all} decay widths studied in this paper vanish in the limit of large $N_c$,
either as $(1/N_c)^2$ or as $1/N_c$. This is true also for decays of exotica, for which the phase space
momentum of the outgoing meson does not vanish in this limit. 

Furthermore we have investigated the large $N_c$ and the Quark Model limits of the decay constants.
In this limit there is a hierarchy of the decay couplings (\ref{eq:gamlimit}): decays of regular baryons are not suppressed, 
pentaquark decay coupling to anti-triplet is suppressed by $1/N_c$, whereas for the sextet the pertinent
coupling vanishes.

\section*{Appendix}
In this Appendix we briefly sketch techniques used to calculate SU(3) Clebsch-Gordan
coefficients for large representations (\ref{eq:reps}). The following Clebsch-Gordan series are relevant
for the decay widths discussed in this paper:
\begin{eqnarray}
(1,1)\otimes\overset{"{\overline{\bf 3}}"}{\overbrace{(0,q+1)}}  &
=&\overset{"\overline{{\bf 15}}"}{\underset{Y_{0}+1,1/2}{\underbrace
{\overbrace{(1,q+2)}}}}\oplus\underset{Y_{0},1}{\overset{"{\bf 6}%
"}{\underbrace{\overbrace{(2,q)}}}}\oplus\,\underset{Y_{0},0}{\overset
{"{\overline{\bf 3}}"}{\underbrace{\overbrace{(0,q+1)}}}} \nonumber \\
&&\oplus
\overset{\text{spurious}}{\underset{Y_{0}-1,1/2}{\underbrace{\overbrace
{(1,q-1)}}}}\; ,
\nonumber\\
(1,1)\otimes \overset{"{\bf 6}"}{\overbrace{(2,q)}}  &  =&\underset
{Y_{0}+1,3/2}{\overset{"{\bf 24}"}{\underbrace{\overbrace{(3,q+1)}}}%
}\oplus\overset{"\overline{{\bf 15}}"}{\underset{Y_{0}+1,1/2}%
{\underbrace{\overbrace{(1,q+2)}}}}\nonumber
\end{eqnarray}
\begin{eqnarray}
& & \oplus\overset{\text{spurious}_{1}}{\underset{Y_{0},2}{\overbrace
{\underbrace{(4,q-1)}}}}\oplus2\,\underset{Y_{0},1}{\overset{"{\bf 6}%
"}{\underbrace{\overbrace{(2,q)}}}}\oplus\underset{Y_{0},0}{\overset
{"{\overline{\bf 3}}"}{\underbrace{\overbrace{(0,q+1)}}}}\nonumber\\
& & \oplus\overset{\text{spurious}_{2}}{\underset{Y_{0}-1,3/2}{\underbrace
{\overbrace{(3,q-2)}}}}\oplus\overset{\text{spurious}_{3}}{\underset
{Y_{0}-1,1/2}{\underbrace{\overbrace{(1,q-1)}}}}\,.
\label{eq:CGseries}
\end{eqnarray}
Labels in quotation marks above  representation labels $(p,q)$ correspond
to the $N_c=3$ limit for these representations, representations that are not
present  for $N_c=3$  are denoted as {\em spurious}. Labels below correspond
to the hypercharge and isospin of the highest weight in a given representation,
with $Y_0=(N_c-1)/3$.

The construction proceeds by starting from the highest weight of the largest
representation in (\ref{eq:CGseries}), for which the SU(3) Clebsch-Gordan coefficient
is 1. Then we apply lowering $I$-spin, $U$-spin and $V$-spin operators to construct
the remaining states in this representation. For explicit form of these operators see {\em e.g.} 
\cite{Behrends}.
Whenever we encounter a state for which an
 orthogonal state exists, we assign it either to  another isospin 
 multiplet in the same representation, or to some lower dimensional
 representation choosing the phases according to de Swart convention \cite{deSwart:1963pdg}. 
 To calculate
 the decay widths we need to construct only $"\overline{\bf 15}"$ and $"{\bf 6}"$ in the
 first series and  $"\overline{\bf 15}"$ in the second. All Clebsch-Gordan coefficients
 have been checked numerically for a few fixed values of $N_c$ with the numerical
 code of Ref.~\cite{Kaeding:1994xh}.

\begin{acknowledgements}
I would like to thank Maxim Polyakov and Hyun-Chul Kim for collaboration
that initiated this research.
This work was supported by the Polish NCN grant 2017/27/B/ST2/01314.
\end{acknowledgements}



\end{document}